\documentclass[pra,aps,superscriptaddress,twocolumn,letter,nopacs,longbibliography]{revtex4-2}

\usepackage[english]{babel}

\usepackage{amsmath, amssymb}
\usepackage{physics}
\usepackage{graphicx,xcolor}
\usepackage[colorlinks=true, allcolors=blue]{hyperref}
\usepackage{ulem}
\usepackage{subcaption}
\usepackage{mathtools}

\usepackage{amsthm,thmtools,thm-restate}

\usepackage{chngcntr}

\newcommand{\navg}{{\bar n}}
\newcommand{\nstdev}{{\Delta n}}

\newcommand{\rhoinput}{\rho_\alpha}
\newcommand{\rhotarget}{\rho_\delta}

\begin{document}

\title{
Competition of decoherence and quantum speed limits for quantum-gate fidelity \\
in the Jaynes-Cummings model
}

\author{Sagar Silva Pratapsi}
\email{spratapsi@tecnico.ulisboa.pt}
\affiliation{Instituto Superior Técnico, Universidade de Lisboa, Portugal}
\affiliation{Instituto de Telecomunicações, Portugal}

\author{Lorenzo Buffoni}
\email{lorenzo.buffoni@unifi.it}
\affiliation{Department of Physics and Astronomy, University of Florence, 50019 Sesto Fiorentino, Italy}

\author{Stefano Gherardini}
\email{stefano.gherardini@ino.cnr.it}
\affiliation{CNR-INO, Area Science Park, Basovizza, 34149 Trieste, Italy}
\affiliation{LENS, Universit\`a di Firenze, I-50019 Sesto Fiorentino, Italy}
\affiliation{ICTP, Strada Costiera 11, 34151 Trieste, Italy}

\begin{abstract}
Quantum computers are operated by external driving fields, such as lasers, microwaves or transmission lines, that execute logical operations on multi-qubit registers, leaving the system in a pure state. However, the drive and the logical system might become correlated in such a way that, after tracing out the degrees of freedom of the driving field, the output state will not be pure. Previous works have pointed out that the resulting error scales inversely with the energy of the drive, thus imposing a limit on the energy-efficiency of quantum computing. In this study, focusing on the Jaynes-Cummings model, we show how the same scaling can be seen as a consequence of two competing phenomena: the entanglement-induced error, which grows with time, and a minimal time for computation imposed by quantum speed limits. This evidence is made possible by quantifying, at any time, the computation error via the spectral radius associated to the density operator of the logical qubit. Moreover, we also prove that, in order to attain a given target state at a chosen fidelity, it is energetically more efficient to perform a single driven evolution of the logical qubits rather than to split the computation in sub-routines, each operated by a dedicated pulse.
\end{abstract}

\maketitle

%%%%%%%%%%%%%%%%%%%%%%%%%%
\section{Introduction}
\label{sec:introduction}
%%%%%%%%%%%%%%%%%%%%%%%%%%

Quantum computing is the science of performing logical operations using quantum states living in a Hilbert space. These logical operations are carried out by unitary gates, in principle in a fully reversible way~\cite{nielsen_chuang_2010,nature2022editorial,Bravyi2022}.
Real quantum computations, however, are limited in practice by the fidelity of quantum operations~\cite{knill2005quantum,cross2019validating}, a problem that is transversal to several quantum platforms~\cite{monz201114,wright2019benchmarking, arute2019quantum,wu2021strong}.
Quantifying the resources in terms of time, energy and space---required to perform a certain gate within a prescribed fidelity---is a task of paramount importance~\cite{Bravyi2022,GiananiDiagnostics2022,fellousasiani2022optimizing}.
We can thus ask, both from a fundamental and operational point of view, whether the resources in realizing quantum gates with a certain fidelity can be bounded.

Recently, some theoretical~\cite{banacloche_minimum_energy, deffner2021energetic,kiely2022classical, aifer2022quantum, ikonen2017energy} and experimental~\cite{cimini2020experimental,stevens2022energetics} studies have put the effort in understanding the resources needed to perform a quantum logical operation by means of an external drive~\cite{Kohler2005driven}. It is known that the entanglement between a driving system, like a laser or an electrical transmission line, and an information-storing logical system, a register of qubits, leads to a fidelity loss when tracing out the drive's Degrees Of Freedom (\textsc{dof}). Previous works have quantified qubit-drive entanglement under a Jaynes-Cummings interaction~\cite{van_enk_classical_2001}, as well as the fidelity loss that it entails~\cite{banacloche_laser_fields}. Others have studied the error arising in two-qubit gates~\cite{banacloche_minimum_energy}, and have investigated whether efficient driving protocols exist~\cite{gea2008quantum}.
Moreover, M.~Ozawa proved that conservation laws also imply fidelity losses~\cite{ozawa_conservative_2002}, which has subsequently been investigated for general unitaries and CPTP maps~\cite{tajima2018uncertainty, PhysRevResearch.2.043374, tajima2021universal, tajima2022universal}.
These results impose  some minimal energy requirements for quantum computation; for example in \cite{banacloche_minimum_energy,banacloche_laser_fields,gea2008quantum}, it has been determined that the computation error scales as $1/\navg$, with $\navg$ being the initial average number of photons (or even phonons) in the drive.
This scaling is important as it can have consequences to energy-efficient computation~\cite{gea2003comparison, moutinho2022quantum, pratapsi2022classical}.

In this work, we (re-)derive and interpret the $1/\navg$ scaling from fundamental principles in the Jaynes-Cummings model. We do so by showing that such a scaling is a consequence of two competing phenomena.
On the one hand, the drive-qubit entanglement induces a minimal error that grows with time. We show that this fundamental error is independent of any target logical state, as it can be quantified via the spectral radius of the density operator of the logical qubit (Section III).
On the other hand, Quantum Speed Limits (QSLs)~\cite{DeffnerPRL2013,DeffnerReview2017} impose a minimal evolution time to attain the target state. The balance between the entanglement error, growing with time, and the minimal time imposed to the dynamics results precisely in a $1 / \navg$ scaling (Section IV).

We also study the error induced by the composition of several quantum gates (Section V). We show that, in order to reach a given target state with a fixed fidelity, it is energetically favourable to use a ``single-pulse'' evolution, instead of splitting the computation in sub-routines, each operated by a dedicated pulse.

%%%%%%%%%%%%%%%%%%%%%%%%%
\section{Driven Quantum Systems}
%%%%%%%%%%%%%%%%%%%%%%%%%
Let us consider a driven quantum system by which we aim to perform logical operations. The drive is modelled as an auxiliary quantum system, \textit{e.g.}, a laser field or an electrical transmission line, such that the logical system and the drive constitute a bipartite quantum system, with local Hilbert spaces $\mathcal{H}_\text{logical}$ and $\mathcal{H}_\text{drive}$ respectively. The state of the bipartite system is denoted by $\ket{\Psi(t)} \in \mathcal{H}_\text{drive} \otimes \mathcal{H}_\text{logical}$. Thus, at any time $t$, the logical state is described by the reduced density operator $\rho(t)$ that is obtained by tracing out the \textsc{dof} of the drive:
\begin{equation}
    \rho(t)
    = \tr_\mathrm{drive} \ketbra{\Psi(t)}
    =: \mathcal{E}_t[\rho(0)] \,,
    \label{eq:effective_quantum_channel}
\end{equation}
where $\mathcal{E}_t[\cdot]$ is a Completely Positive Trace Preserving (CPTP) quantum map~\cite{Breuer2002}. In quantum information theory, $\mathcal{E}_t[\cdot]$ is commonly denoted as a quantum channel~\cite{CarusoRMP2014}.

In the remainder of the text, we will assume that the initial logical state is pure so that $\rho(0) = \rhoinput := \ketbra{\alpha},$ for some state $\ket{\alpha}$, in the density operator formalism.

%%%%%%%%%%%%%%%%%%%%%%%%%
\subsection{Jaynes-Cummings Model}
%%%%%%%%%%%%%%%%%%%%%%%%%

Throughout the paper, we will study the paradigmatic example of the Jaynes-Cummings (JC) model. The Hamiltonian of the bipartite system is given by
\begin{equation}\label{eq:JC_hamiltonian}
H = \hbar \omega \, b^\dagger b
+ \hbar \omega \, \ell^\dagger \ell
+ \hbar g i (b\ell^\dagger - b^\dagger \ell),
\end{equation}
where $b,\ell$ are the ladder operators for the drive and the logical system, respectively.
For the states of the bipartite system, we adopt the notation $\ket{n,m} := \ket{n}_\mathrm{drive} \otimes \ket{m}_\mathrm{logical}$.

In an appropriate rotating frame, the terms proportional to $\hbar \omega$ disappear, and we are left with the interaction Hamiltonian
\begin{equation*}
    H_\mathrm{int}
    =
    \hbar g i (b\ell^\dagger - b^\dagger \ell)\,.
\end{equation*}
From here on, we will work in such a frame.
Note that $H_\mathrm{int}$ fixes the rotation axis (here, the $y$-axis) of the logical qubit along with the interaction through which the driving fields acts. This choice, however, is arbitrary and is connected to the choice of an arbitrary phase of the driving field. The analysis that follows does not take into account any phase-related arguments but just the average number of photons injected by the drive to the logical system.

We assume that the drive starts in some initial state $\ket{\varphi_\mathrm{dr}(0)} = \sum_{n=0}^\infty b_n \ket{n_{\rm dr}}$, which we can control.
The levels $\ket{n, 0}$ and $\ket{n-1, 1}$ span an invariant subspace of $H$, with respect to which it is convenient to solve the Schr\"{o}dinger equation that models the whole evolution of the JC model. In this subspace, the dynamics oscillate between $\ket{n, 0}$ and $\ket{n-1, 1}$ at the frequency $\omega_n := g\sqrt{n}$.
It is then a straightforward matter to calculate $\ket{\Psi(t)}$. The logical density operator $\rho(t)$ can be fully described as a linear combination of the operators $\mathcal E_{ij} := \mathcal E_{t}[\ketbra{i}{j}]$ at any time $t$.
Each $\mathcal E_{ij}$, in turn, is provided by the average of the $2\times 2$ operators $F_{ij}$ (for the sake of completeness we recall their explicit expression in Appendix~\ref{appendix:JC_evolution}) over the energy levels distribution of the initial drive state. Formally, it means that $\mathcal E_{ij} = \sum_{n=0}^\infty \abs{b_n}^2 F_{ij}(n) = \mathbb{E}_{\rm dr}[ F_{ij} ]$, with $\mathbb{E}_{\rm dr}[\cdot]$ denoting the average over the set of drive coefficients $\{b_n\}$. Thus, $\mathcal E_{ij}$ is returned by a series that, in the general case, one is not able to solve.

The operators $F_{ij}(n)$ can be expanded in Taylor series around $\navg := \expval{b^\dagger b}$, i.e., 
\begin{equation*}
    F_{ij}(n)
    = F_{ij}(\navg) + \left.F_{ij}'\right|_{n=\navg}(n - \navg) + \frac{1}{2}\left.F_{ij}''\right|_{n=\navg}(n-\navg)^2 + \cdots 
\end{equation*}
with $(\cdot)'$ and $(\cdot)''$ denoting, respectively the 1st and 2nd derivatives of $(\cdot)$ with respect to $n$. As a result, by recalling that $\mathbb E_{\rm dr}[n-\navg] = 0$, one gets
\begin{equation}\label{eq:expval_approximation}
    \mathcal E_{ij} =
    \mathbb E_{\rm dr}[F_{ij}] \approx F_{ij}(\navg) + \frac{1}{2}\left.F_{ij}''\right|_{n=\navg}\nstdev^2 \,,    
\end{equation}
where $\nstdev^2 := \mathbb E_{\rm dr}[(n-\navg)^2]$ is the variance of $n$. The validity of this second-order expansion is assured by taking the initial state of the drive so that $\nstdev \leq \navg$, thus meaning that the initial distribution of the driving field is mostly concentrated around the mean value (e.g., mean number of photons) $\navg$.

In order to provide results with an operational meaning, we are going to consider two different cases for the initial state of the drive: a coherent state (thus, with Poisson distribution) with mean $\navg$, and a binomial-distribution state with mean $\navg$ and standard deviation $\Delta n$. Both the Poisson and binomial distributions are concentrated around the mean value $\navg$, and the coefficients $b_n$ of the initial state of the drive are respectively equal to
\begin{equation}\label{eq:distributions_photons}
    b_n^\textrm{Pois.}
    = \frac{1}{e^{\navg/2}} \frac{\navg^{n/2}}{\sqrt{n!}}
    \quad \text{and} \quad
    b_n^\textrm{Bin.}
    = \frac{1}{2^{N/2}} \sqrt{ \binom{N}{k_n} } \,.
\end{equation}
In Eq.~(\ref{eq:distributions_photons}), $N := 2\Delta n^2$ is the width of the binomial distribution and $k_n := n - (\navg - N)$. At any time $t$, we can estimate each $\mathcal E_{ij}$ in second-order approximation as given by Eq.~\eqref{eq:expval_approximation}.

%%%%%%%%%%%%%%%%%%%%%%%%%
\section{Entanglement-induced error}
\label{section:entanglement_induced_error}
%%%%%%%%%%%%%%%%%%%%%%%%%

As stated in the Introduction, the aim of this section is to provide a bound on the computation error that is made by the logical system due to the presence of `residual' drive-qubit entanglement, still present after tracing out the \textsc{dof} of the drive. Such an error induced by the drive-qubit entanglement is independent of any target logical state.

In general, the fidelity between the output logical state $\mathcal E_t[\rhoinput]$ and some target density operator $\rhotarget := \ketbra{\delta}$ is
\begin{equation}\label{eq:QSL-general}
    F(\rho(t),\rhotarget)
    = \expval{\mathcal E_t \left[ \ketbra{\alpha} \right]}{\delta}.
\end{equation}
In fact, since the target density operator $\rhotarget$ represents a pure state, $F(\rho(t),\rhotarget)$ corresponds to the Uhlmann fidelity~\cite{jozsa1994fidelity,miszczak2008sub}. The latter is inversely proportional to the Bures angle that separates $\rhotarget$ to $\mathcal E_t [\ketbra{\alpha}]$, in the space of density operators pertaining to the quantum logical system. On the other hand, $\mathcal E_t$ contains information about the evolution of the dynamics developed by the logical system, and about quantities that are representative of the amount of energy that is injected by the drive.

First, we introduce some quantities that provide fundamental bounds to the fidelity $F(\rho(t),\rhotarget)$ that is associated with the target transformation $|\alpha\rangle \rightarrow |\delta\rangle$ with $|\alpha\rangle$, $|\delta\rangle$ pure states. 
These bounds do not depend on the specific choice of the initial and target states $|\alpha\rangle$ and $|\delta\rangle$. However, we can characterize the intrinsic error due to the loss of purity of the reduced states of both the logical system and the drive. In addition, we manage to obtain predictions that impart valuable insights into the best-case scenario for the computation fidelity, thus offering a glimpse into the potential upper limits of attainable accuracy due to fundamental limitations. Furthermore, by removing the knowledge of the target state $|\delta\rangle$, we obtain a quantity that can be measured directly from $\rho(t)$, and possibly scaled to large quantum systems where the knowledge of the target state is computationally unfeasible.

%%%%%%%%%%%%%%%%%%%%%%%%%
\subsection{State eigenfidelity}
%%%%%%%%%%%%%%%%%%%%%%%%%

The first quantity we introduce is the state eigenfidelity. For this purpose, let us consider the following Proposition (the proof is in Appendix~\ref{appendix:eigenfidelity}):
\begin{restatable}{prop}{propone}\label{prop:closest_state_fidelity}
    The closest \textit{pure} states to a generic density operator $\varrho$ are the eigenstates (or eigenstate, in the non-degenerate case) $\ket \psi$ corresponding to the largest eigenvalue $r(\varrho)$ of $\varrho$, also denoted as its \textit{spectral radius}. Formally,
    \begin{equation}
    F(\varrho,\rho_{\varphi}) \leq r(\varrho)
    \end{equation}
    for any pure state $\rho_{\varphi} := |\varphi\rangle\!\langle\varphi|$, and $F(\varrho,\rho_{\varphi}) = r(\varrho)$ if and only if
    $\varrho \ket{\varphi} = r(\varrho) \ket{\varphi}$.
\end{restatable}
Hence, the error done at any time $t$ while performing the transformation $|\alpha\rangle \rightarrow |\delta\rangle$ through the quantum channel $\mathcal E_t$---truly implemented in a real context---is at least equal to $1 - r\left( \mathcal E_t [\ketbra{\alpha}] \right)$. For the fidelity $F(\varrho,\rho_{\varphi})$, the spectral radius $r(\varrho)$ is a bound that stems from the lack of purity of $\varrho$. For this reason, we denote $r(\varrho)$ as the \textit{eigenfidelity} of $\varrho$, and $\epsilon(\varrho) := 1 - r(\varrho)$ the corresponding \textit{eigenerror}.
Although this definition suffices for our purposes, we note that the eigenfidelity cannot be taken in the general case as a guarantee of how good is a given quantum gate, but as a limit of how good a given quantum gate could be in the best-case scenario. In this regard, see Refs.~\cite{gea1990collapse,gea1991atom} for an example of a non-unitary process where the eigenfidelity is nonetheless high.

In the following, we are going to provide lower and upper bounds for the eigenerror $\epsilon(\varrho)$.
In doing this, we start by considering the Schatten $p$-norm of a linear operator. For a density operator $\varrho$, the Schatten $p$-norm is defined as $\norm{\varrho}_p := (f_1^p + \cdots + f_n^p)^{1/p}=(\tr\varrho^p)^{1/p}$, where $f_i$ is the $i$-th eigenvalue of $\varrho$ and $p$ is a positive real number.
We thus obtain:
\begin{restatable}{prop}{proptwo}\label{prop:Schatten_norm}
    For any $p > 1,$ spectral radius of a density operator $\varrho$ is bounded from below and above as
    \begin{equation}\label{eq:bound_with_norm_p}
        \norm{\varrho}_p^q \leq r(\varrho) \leq \norm{\varrho}_p \,,
    \end{equation}
    where $q$ is such that $1/q + 1/p = 1.$
\end{restatable}
The proof is in the Appendix~\ref{appendix:eigenfidelity}, and exploits the fact that the spectral radius of a generic density operator $\varrho$ is a lower bound of any matrix norm (expressed in terms of $p$) of $\varrho$. Moreover, from the proof, one can also deduce that the bounds in Proposition~\ref{prop:Schatten_norm} become tighter and converge to the eigenfidelity $r(\varrho)$ as $p$ goes to infinity. For applications in quantum technologies, just the Frobenius norm $\norm{\varrho}_F := \sqrt{\tr \varrho^2} = \norm{\varrho}_2$ suffices. In this way, by making the first-order expansion in Taylor series of the upper bound in \eqref{eq:bound_with_norm_p} around $\tr\varrho^2 = 1$, for $r(\varrho)$ we get the following bounds with an \textit{operational} meaning:
\begin{restatable}{thm}{thmone}\label{prop:fidelity_upper_bound}
    The eigenfidelity $r(\varrho)$ and eigenerror $\epsilon(\varrho)$ of $\varrho$ are bounded from below and above as
    \begin{equation}\label{eq:Prop_2}
        \gamma
        \leq r(\varrho)
        \leq \frac{1 + \gamma}{2}
        \quad\text{and}\quad
        \frac{S_L}{2}
        \leq \epsilon(\varrho)
        \leq S_L \,,
    \end{equation}
    where $\gamma := \tr \varrho^2$ and $S_L := 1 - \gamma$ denote, respectively, the purity and the linear entropy of $\varrho$.
\end{restatable}
Thus, if $\varrho$ represents a pure quantum state, then $r(\varrho)=\gamma=1$. Instead, if $\varrho$ is a maximally mixed state, then $r(\varrho)=\gamma=1/d$, with $d$ dimension of the quantum logical system. In these two cases, the bounds of $r(\varrho)$ in Eq.~\eqref{eq:Prop_2} become trivial. However, in full generality, the upper bound $(1+\gamma)/2$ of the eigenfidelity $r(\varrho)$ is as much tight as the error in performing the desired logical operation is small. We can thus affirm that $r(\varrho) \sim (1 + \gamma)/2$ becomes a good approximation when $\gamma$ is close to 1, i.e., when $\varrho$ is nearly a pure state, as all the higher-order corrections $(1-\gamma)^2, (1-\gamma)^3, \ldots$ to $r(\varrho)$ are vanishing. Clearly, these considerations also hold for the eigenerror $\epsilon(\varrho)$ but with a reversed meaning. 

We conclude our analysis by providing a different interpretation of the eigenfidelity that links to a series of recent works dealing with the purification of states~\cite{ticozzi2014quantum} and the effects entailed by the presence of an external environment~\cite{guryanova2020ideal,BuffoniPRL2022,mohammady2023quantum}. Let us thus take the definition of \textit{ergotropy}~\cite{allahverdyan2004maximal}, whereby any density operator $\varrho = \sum_{i} s_i |s_i\rangle\!\langle s_i|$ can be unitarily mapped into the corresponding \textit{passive state} $\sigma := \sum_{i} s_i |e_i\rangle\!\langle e_i|$, where $\{ \ket{e_i} \}$ is the energy basis of the quantum system with Hamiltonian $H=\sum_{i}e_{i}|e_i\rangle\!\langle e_i|$ such that $e_1 < e_2 < \ldots < e_d$. By construction, a passive state is diagonal in $\{ \ket{e_i} \}$, and with $s_1 > s_2 > \ldots > s_d$. From these definitions we can see that $s_1$ is exactly the eigenfidelity. Consequently, we can state that \textit{different density operators $\varrho$ that share the same passive state have the same eigenfidelity}. 

%%%%%%%%%%%%%%%%%%%%%%%%%
\subsection{Channel eigenfidelity}
%%%%%%%%%%%%%%%%%%%%%%%%%

Let us now apply the concept of eigenfidelity to bound the performance of a quantum channel $\mathcal{E}_t$ over a complete set of input and target pure states. Thus, for any time $t$, we average the eigenfidelity $r(\mathcal{E}_t[\rho_{\alpha}])$ over any initial pure states $\rho_{\alpha}=|\alpha\rangle\!\langle\alpha|$ sampled in accordance with the (normalized) Haar's measure $\mu(\rho_{\alpha})$, i.e., 
\begin{equation}
\bar r \left( \mathcal{E}_t \right) := \int \dd \mu( \rho_{\alpha} ) \, r( \mathcal E_t[ \rho_{\alpha} ] ) \,.
\end{equation}
We denote $\bar r \left( \mathcal{E}_t \right)$ as the \textit{channel eigenfidelity} that is \textit{independent} of both the initial and final pure quantum states. The quantity $1-\bar r \left( \mathcal{E}_t \right)$ corresponds to the error committed by a quantum channel while approximately operating a desired logical quantum operation; we thus call it as the \textit{channel eigenerror}. Such an error is proportional to the impurity of $\mathcal{E}_t[\rho_{\alpha}]$ averaged over all possible $\rho_{\alpha}$. In this regard, thanks to Theorem~\ref{prop:fidelity_upper_bound}, also the channel eigenfidelity can be bounded from below and above:
\begin{restatable}{cor}{cor_one}\label{cor:avg_eigenfidelity_bounds}
    The channel eigenfidelity $\bar{r}(\mathcal E_t)$ of a quantum channel $\mathcal E_t$, for any time $t$, is bounded as
    \begin{equation}
        \label{eq:avg_eigenfidelity_bounds}
        \bar \gamma_t \leq \bar r(\mathcal E_t) \leq \frac{1 + \bar \gamma_t}{2}\,,
    \end{equation}
    where $\bar \gamma_t$ is the average of the purity $\tr \mathcal E_{t}[\rho_{\alpha}]^2$ over all input pure states $\rho_{\alpha}$.
\end{restatable}
\noindent
Moreover,
\begin{restatable}{prop}{qubitpurity}\label{prop3_average_purity}
For qubits, $\bar \gamma_t$ has a closed-form expression:
\begin{equation}\label{eq:best_case_fidelity}
\bar \gamma_t = \frac{1}{3} \tr \left(
     \mathcal E_{00}^2
     + \mathcal E_{00} \, \mathcal E_{11}
     + \mathcal E_{11}^2 
     + \mathcal E_{01} \, \mathcal E_{10}
    \right),
\end{equation}
where, we recall, $\mathcal E_{ij} = \mathcal E_t\left[\ketbra{i}{j}\right]$ with $i,j=0,1$.
\end{restatable}
While the derivation of the Corollary \ref{cor:avg_eigenfidelity_bounds} is not needed being a direct application of Theorem~\ref{prop:fidelity_upper_bound}, the proof of Proposition \ref{prop3_average_purity} is in Appendix~\ref{appendix:eigenfidelity}. 

%%%%%%%%%%%%%%%%%%%%%%%%%
\subsection{Channel eigenerror of the Jaynes-Cummings evolution}\label{sec:JC_Ham}
%%%%%%%%%%%%%%%%%%%%%%%%%

In the Jaynes-Cummings setting, the channel eigenerror $\bar \epsilon_t = 1 - \bar{r}(\mathcal E_t)$ can be studied in the asymptotic limit
$\navg \to \infty$ with the constraint that the \textit{Fano factor} $s := \nstdev^2 / \navg$ is kept \textit{constant}. Specifically, one can analytically determine that 
\begin{equation}\label{eq:eigenerrors_JC_model}
    \bar \epsilon_t
    \geq \frac{\bar S_L}{2}
    \sim
    \begin{dcases}
        \frac{\tau^2 + \sin(\tau)^2}{6 \navg}
        & \mathrm{(Poisson)}
        \\
        \frac{\tau^2}{6} \frac{\Delta n^2}{\navg^2}
        + \frac{\sin(\tau)^2}{6 \, \Delta n^2}
        & \mathrm{(Binomial)}
    \end{dcases}
\end{equation}
by having the initial state of the drive in
a Poisson or binomial distribution, respectively. In Eq.~(\ref{eq:eigenerrors_JC_model}), $\tau := \omega_\navg t = g\sqrt{\navg}\,t$ denotes the reduced interaction time, expressed in radians. These bounds follow from the explicit computation of Eq.~(\ref{eq:best_case_fidelity}) applied to the JC model under the approximation of Eq.~(\ref{eq:expval_approximation}), which we have carried out via the symbolic software SageMath~\cite{sagemath}. 
As a remark, it is worth observing that, since $s$ is kept constant, the assumption that the distribution of the drive state is concentrated around its mean remains valid in the $n \to \infty$ limit, in line with the Taylor expansion in~\eqref{eq:expval_approximation}. We also note that the approximation~\eqref{eq:eigenerrors_JC_model} holds for short times only due to the presence of the quadratic term $\propto\tau^2$.
Moreover, $\nstdev^2 = \navg$ for the Poisson distribution; so, making this substitution in~\eqref{eq:eigenerrors_JC_model}, the bound for the binomial distribution reduces to the one of the Poisson distribution, such that both expressions are consistent.

\begin{figure*}[t]
    \centering
    \includegraphics[width=0.99\linewidth]{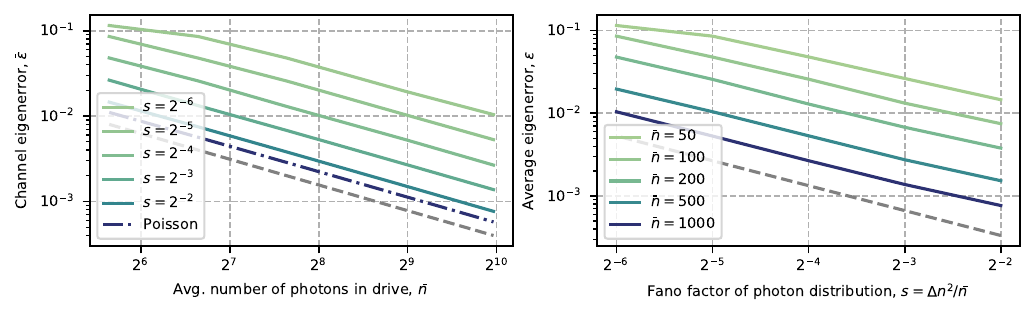}
    \caption{Channel eigenerror of the quantum logical system operated by the JC model with Hamiltonian (\ref{eq:JC_hamiltonian}), for 
    the reduced interaction time $\tau = \pi/2$.
    \textit{Left panel}: Eigenerror as a function of the average number of photons in the initial state of the drive, for different values of $s \in [0.02, 0.25]$ (colored solid lines). The dash-dot line corresponds to a Poisson distribution for the drive state, which has $s=1.$
    \textit{Right panel}: Eigenerror as a function of the Fano factor $\nstdev^{2}/\navg$ of the initial state of the drive, for different values of $\navg \in [50, 1000]$
    In the panels, the grey dashed line denotes the reference scaling of $1/\navg$ (left panel) and $1/s$ (right panel), respectively.}
    \label{fig:eigenerror_scaling}
\end{figure*}

We illustrate in Fig.~\ref{fig:eigenerror_scaling} the scaling of the channel eigenerror $\bar \epsilon_t$ as provided by Eq.~\eqref{eq:eigenerrors_JC_model} as a function of $\navg$ and $s = \nstdev^2 / \navg$, for the reduced interaction time $\tau=\pi/2$. On the left panel, we see that the channel eigenerror decreases as $1/\navg$ by taking constant the Fano factor $s$, for both the initial states given by a Poisson and binomial distribution. On the right panel, instead, $\bar \epsilon_t$ decreases as $1/s$ for constant $\navg$. Notice that the latter behaviour can be only inferred from the binomial distribution, being $s=1$ for the Poisson one.
Interestingly, Eq.~\eqref{eq:eigenerrors_JC_model} suggests that $s = \sin(\tau) / \tau$ minimizes the error for a given $\tau$, in line with previous observations using squeezed states~\cite{ikonen2017energy}, although---strictly speaking---the binomial distribution does not allow for values of $s$ larger than $1/2$.

The bounds in Eq.~\eqref{eq:eigenerrors_JC_model} capture the interesting worst-case scenario of $\Delta n \to 0$ for the JC model. Consider, for example, the initial bipartite state $\ket{N,0}$, and require that at $\tau = \pi/4$ the logical state is $(\ket 0 + \ket 1)/\sqrt{2}$, as a $\pi/2$-pulse were applied. In such a case, the density operator of the logical system is the maximally mixed state, no matter how high $N$ is, as long as $\Delta n = 0$. Hence, the error of the computation turns out to be the worst possible one. Likewise, the error bounds in Eq.~\eqref{eq:eigenerrors_JC_model} (correctly) diverge when $\nstdev$ goes to zero.

The bounds in Eq.~\eqref{eq:eigenerrors_JC_model} reproduce previous results for the error done by the $X$ and $\sqrt{X}$ gates~\cite{banacloche_laser_fields}, 
where the error committed in realizing a driven single-qubit operation operated by a JC Hamiltonian scales as $\epsilon \gtrsim \left(\nstdev/\navg\right)^2$. As noted in \cite{banacloche_laser_fields}, this has implications for the energetics of quantum computation: in order to lower the eigenerror, we need to increase $\navg$, which means using a driving system with more average injected energy. Moreover, the scaling in Eq.~\eqref{eq:eigenerrors_JC_model} has also been found for two-qubit gates \cite{banacloche_minimum_energy}, and for the inaccuracy of generating entanglement from a bipartite system under the JC Hamiltonian (\ref{eq:JC_hamiltonian})~\cite{van_enk_classical_2001}.

In summary, the bounds of the channel eigenerror $\bar\epsilon_t$ contain the same characteristic features of other scaling laws previously determined in the literature. Such bounds mainly depend on the intensity and the shape of the interaction terms between the drive and the quantum logical system, according to the JC model.
The error figures we have determined capture an ``intrinsic'' error that is due to the average output state not being pure, and the resulting bounds are more accurate as long as the reduced interaction time $\tau$ is small.
However, it is worth stressing that a low channel eigenerror is not a guarantee of a quantum operation with high fidelity. Conversely, a high channel eigenerror entails that the implemented quantum operation has a low fidelity.

In the next section, we are going to show a clear connection of the channel eigenerror with quantum speed limits. This will allow us to understand how the energetic cost of a quantum logical computation is fundamentally based on a trade-off between two factors: the bare error that remains after making the partial trace with respect to the drive state, and users' requirements about the minimal time needed to carry out the computation.

%%%%%%%%%%%%%%%%%%%%%%%%%
\section{Quantum Speed Limits and Energetics of Quantum Computation}
%%%%%%%%%%%%%%%%%%%%%%%%%

In the previous section, we have evaluated the error that a quantum channel intrinsically introduces in the attempt to return as output a pure state that is the solution of a given quantum computation.
In the limit of a small computation time, our analysis does not take into account whether the implemented quantum operation returns the target output state. For the logical operation operated by the Jaynes-Cummings dynamics, we have shown that in the short timescales before the state collapse, the eigenerror grows as a function of the elapsed time $t$.

However, in practice, one aims to understand what are the constraints imposed by quantum mechanics to the scaling of the computation error in the attempt to apply a certain unitary gate within a fixed time interval (possibly chosen by the user). We would also like to determine the energetic cost associated to respecting such constraints. We are going to introduce these additional notions by appealing to quantum speed limits.

The Mandelstamm-Tamm (MT) and Margolus-Levitin (ML) QSLs~\cite{DeffnerReview2017} provide restrictions on the minimal time needed for a quantum system to perform unitary evolutions. Specifically, the time $t$ required for a $\vartheta$ rotation in the Hilbert space of the system is bounded, respectively, as
\begin{equation}\label{eq:qsl}
    t \geq t_\text{MT} := \frac{\hbar \vartheta}{\Delta H_\mathrm{int}}
    \quad\text{and}\quad
    t \geq t_\text{ML} := \frac{\hbar \vartheta}{\expval{H_\mathrm{int}}}\,,
\end{equation}
where $H_\mathrm{int}$ is the Hamiltonian of the quantum system, $\langle X\rangle$ denotes the expectation value of $X$, and $\Delta X^2 := \expval{X^2}-\expval{X}^2$. All the averages $\langle\cdot\rangle$ are performed with respect to the initial state of the bipartite quantum system.
Note that, since the QSLs in Eq.~(\ref{eq:qsl}) are defined for unitary evolutions, we can apply them only to the whole bipartite system. Thus, the quantity $\vartheta$ refers to angles in the Hilbert space comprising both the logical and the driving system, and the QSLs depend on the bipartite interaction Hamiltonian $H_\mathrm{int}$.

However, we are interested in the evolution of the logical (sub-)system, and thus to link the QSLs to such reduced dynamics.
For this purpose, suppose we aim to bring a pure logical state $\ket a$ to another pure logical state $\ket b$, such that $\abs{\braket{b}{a}} =: \cos\theta$, with $\theta\in [0,\pi/2]$ in order to avoid issues with the sign of the cosine function.
Hence, the bipartite system consisting of the logical qubit and the drive should go from the initial pure state $\ket \psi = \ket {\psi_a} \ket a$ to a final pure state $\ket{\psi'} = \ket{\psi_b}\ket b$.
The angle $\vartheta$ between the initial and final states of the bipartite system fulfils the inequality
\begin{equation}
    \cos \vartheta
    = \abs{\braket{\psi'}{\psi}}
    = \abs{\braket{\psi_b}{\psi_a}} \cos \theta
      \leq \cos\theta \,.
\end{equation}
This implies that $\vartheta \geq \theta$, since $\theta$, $\vartheta \in [0,\pi/2]$. 
As a result, the angle $\theta$ and the QSLs for $t_\text{MT}$, $t_\text{ML}$ find direct connection through the inequality
\begin{equation}\label{eq:qsl_logical}
    t
    \geq
    \mathrm{max}\left\{
    \frac{\hbar \theta}{\Delta H_\mathrm{int}}
    ,
    \frac{\hbar \theta}{\expval{H_\mathrm{int}}}
    \right\},
\end{equation}
that establishes a geometric constraint on the driving of the logical system, which fixes the whole energy resources at disposal.

%%%%%%%%%%%%%%%%%%%%%%%%%
\subsection{Jaynes-Cummings Hamiltonian}
%%%%%%%%%%%%%%%%%%%%%%%%%

For the sake of clarity we provide more analytical arguments still using the JC model. In doing this, let us suppose the drive is initialized in a coherent Poisson state $\ket{\psi_a}$ with average photon number $\navg$.
In such a case, we have $\expval {H_\mathrm{int}} \approx \Delta H_\mathrm{int} \sim \hbar\,\omega_{\navg},$ with $\omega_\navg = g \sqrt{\navg}$. Moreover, in the large $\navg$ limit, the Poisson and binomial distributions tend to coincide when $\Delta n = \sqrt{\bar n}$.
Therefore, both the MT and ML QSLs~\eqref{eq:qsl_logical} impose the following lower bound for the gate operation time (driving field included):
\begin{equation}
    t \geq \frac{\theta}{\omega_{\navg}}
    = \frac{\theta}{g \sqrt{\navg}}\,.
    \label{eq:qsl-time-scaling}
\end{equation}

On the other hand, still for a Poissonian drive, Eq.~(\ref{eq:eigenerrors_JC_model}) predicts a minimal average eigenerror that grows with time that explicitly reads as
\begin{equation}\label{eq:error-time-scaling}
\bar \epsilon_t \geq
\frac{g^2 t^2}{6} + \frac{\sin^2(\sqrt{\navg} g t)}{6 \navg}\,.    
\end{equation}
We note that the QSL times are of the order of a single Rabi oscillation (thus before a collapse in the reduced logical state of the JC dynamics), where this scaling is valid.

Putting both inequalities together, we determine that to perform a rotation $\theta$ on the logical qubit using a JC interaction, the lower bound of the channel eigenerror entailed by the QSL scales as
\begin{equation}\label{eq:bar_epsilon_theta_1}
    \bar\epsilon_\theta \gtrsim \frac{\theta^2 + \sin^2(\theta)}{6\navg}
\end{equation}
that, for small angles ($\sin \theta \sim \theta$), becomes 
\begin{equation}\label{eq:bar_epsilon_theta_2}
   \bar\epsilon_\theta \sim \frac{ \theta^2 }{ 3\navg }\,. 
\end{equation}
Note that the average number of driving photons $\navg$ is proportional to the initial energy put into the system by the driving system, since the Hamiltonian of the latter is $H_\mathrm{drive} = \hbar\omega b^\dagger b$.
Our derivation thus encloses the findings of previous works~\cite{ozawa_conservative_2002, banacloche_minimum_energy} stating the following: if one aims that the average error in performing a single-qubit computation is smaller than $\epsilon$, then a driving pulse with
energy $E = \navg \hbar \omega$
scaling as $1/\epsilon$ has to be employed. This analysis provides us that the energetic requirement to drive a quantum logical system, at least in the Jaynes-Cummings model, comes from the balance of two competing phenomena:
(i) the entanglement-induced error from Eq.~\eqref{eq:error-time-scaling}, increasing with time, that suggests one should use shorter gate times; and (ii) the QSLs of Eq.~\eqref{eq:qsl-time-scaling}, which tell us that shorter gate times come at the cost of increasing the number of photons $\navg$.
This implies that, in the case we considered (i.e., the JC model), the strength of the driving field needs to have enough photons (that is, energy) according to Eqs.~(\ref{eq:bar_epsilon_theta_1})-(\ref{eq:bar_epsilon_theta_2}) in order to make negligible the error due to inefficient use of the laser-qubit entanglement.

%%%%%%%%%%%%%%%%%%%%%%%%%
\section{Quantum gate concatenations}
%%%%%%%%%%%%%%%%%%%%%%%%%

In this section, we generalize the error bounds introduced in the previous sections for the case of $C$ successive applications of a quantum channel $\mathcal{E}_t$, which for simplicity we will call ``shot'' or ``pulse shot''.
As in Eq.~\eqref{eq:effective_quantum_channel}, each shot $\mathcal E_{t}$ results from i) preparing the driving system in some initial pure state with average number of photons $\navg$; ii) letting the bipartite system evolve according to the Jaynes-Cummings evolution for some time $t$; iii) tracing out the \textsc{dof}s of the driving system.
In this section, we are going to quantify the channel eigenerror of a multiple shot evolution, i.e.,
\begin{equation}
    \mathcal E_t^C
    = \underbrace{
        \mathcal E_t
        \circ \mathcal E_t
        \circ \cdots
        \circ \mathcal E_t
        }_\text{$C$ times}\,.
\end{equation}
Physically, this implies to restart the driving state $C$ times, each time introducing $\navg$ new photons into the system and discarding the old ones.

Let $\bar\epsilon_{\tau,C}$ be the average eigenerror of $\mathcal E_t^C$. For simplicity of notation, we use the reduced time $\tau = g\sqrt{\navg} t$. In the large $\navg$ limit, $\tau$ is approximately equal to the rotation angle of the logical state; thus, after $C$ applications of $\mathcal E_t$, we expect the logical state to rotate by the total angle $C\tau$. Our aim is to understand---from first principles---whether it is better to implement fewer times a quantum operation of longer duration, or more times a gate whose duration is shorter.

\begin{figure*}[t]
    \includegraphics{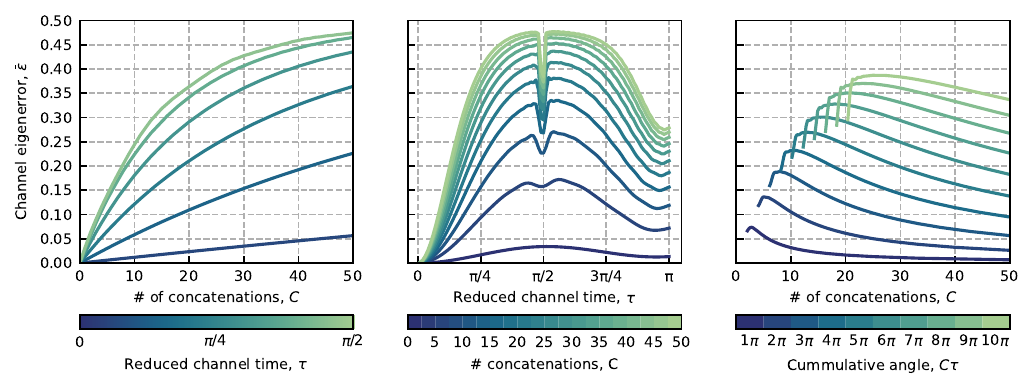}
    \caption{
    Channel eigenerror in concatenating a quantum operation. In the numerical simulations, the concatenated quantum gates are enabled by a JC Hamiltonian, and the drive is initialized in a binomial state with $\navg=25$ and Fano factor $s = \nstdev^2 / \navg = 0.2$.
    \textit{Left panel}: The channel eigenerror grows monotonically with the number of concatenations, increasing faster for greater values of $\tau$.
    \textit{Central panel}: Channel eigenerror as a function of the channel reduced time $\tau$. Each line corresponds to a fixed number $C$ of concatenations. See the main text for the explanation of the behaviour of the channel eigenerror  around $\tau = \pi/2$.
    \textit{Right panel}: Channel eigenerror as a function of the concatenations number. Each line corresponds to a constant cumulative evolution time $C\tau$. For the same cumulative evolution time, we achieve a lower channel eigenerror for the concatenation of a larger number of gates (larger $C$) but with a shorter reduced interaction time $\tau$. However, this comes with a larger energetic cost, see Fig.~\ref{fig:concatenation_energy_efficiency} for details.
    }\label{fig:concatenations}
\end{figure*}

For the purposes of this section, we assume that the driving system is initialized in a state with a binomial distribution, also denoted as `binomial state' for simplicity. In Appendix~\ref{appendix:concatenation}, we prove that the channel eigenerror approximately reads as
\begin{equation}
    \bar\epsilon_{\tau,C}
    \approx 
    \frac{1 - \lambda^{C}}{2}\,,
\end{equation}
where $\lambda$ is the spectral radius of the $3\times 3$ sub-matrix in the representation of $\mathcal{E}$ in terms of the Pauli Transfer Matrix (PTM) formalism. For more details, see the Appendix~\ref{appendix:concatenation} where analytical expressions are provided. This formula is tested with numerical simulations, setting $\navg=25$ and the Fano factor $s = \nstdev^2 / \navg = 0.2$.
Notice that our interest in considering a binomial distribution stems from showing a peculiar behaviour (a dip) of the channel eigenerror when plotted as a function of the reduced interaction time $\tau$.

From the left panel of Fig.~\ref{fig:concatenations}, showing numerical results, one can observe that the channel eigenerror increases monotonically with the number of concatenations $C$, no matter the reduced interaction time $\tau$. Hence, the repeated application of the same gate operated by a short interaction with the driving system seems to reduce the channel eigenerror.

On the other hand, the right panel of Fig.~\ref{fig:concatenations} shows how a greater number of concatenations, for the same total evolution time $C\tau$, leads to a smaller channel eigenerror. In other words, if one wishes to implement a given quantum operation, it is beneficial to break the target quantum operation in many dynamical cycles of a shorter duration $\tau$. However, such division comes with a \textit{higher} energetic cost, since every time a gate is applied, one needs to prepare the initial state of the drive from scratch (thus, with new photons ready to use), such that in total $C\navg\hbar\omega$ of energy is used.

The central panel of Fig.~\ref{fig:concatenations} shows a dip in the channel eigenerror at $\tau=\pi/2$. This behaviour does not occur if the driving system is initialized with a Poisson distribution, and it has to be considered a hallmark of the quantum nature of the initial state of the driving field. In fact, by taking the limit $s \to 0$ (for the case of binomial state) with a value of $\navg$ sufficiently large, the channel eigenerror at $\tau=\pi/2$ is $0.25$ and becomes independent from the number $C$ of concatenations. This dip in the channel eigenerror at $\tau=\pi/2$ can be observed until values of $s$ up to $s=0.2$. Instead, for all the other values of $\tau$, the state of the logical system rapidly approaches the mixed state under performing quantum gate concatenation, and the channel eigenerror generally increases.

As briefly introduced above, one would ask whether, for the same target eigenerror, there is an \textit{energetic advantage} from subdividing a given pulse shot $\mathcal{E}$ into a concatenation of multiple shots of shorter duration.
To answer this question, suppose that we want to rotate the logical state by an angle $\theta$, by having at our disposal a total of $\navg$ photons. To do this, we can either perform a single shot evolution of the angle $\theta$---via a driving field using all the $\navg$ available photons---or we can implement $C$ shots, each using $\navg/C$ photons and each rotating the logical state by the angle $\theta/C$. Note that, in the latter case, each pulse must be applied for the time $t = \theta / g\sqrt{C \navg}$.

\begin{figure}[t]        
    \includegraphics{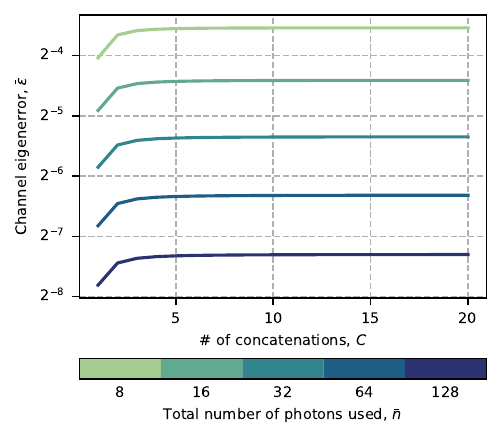}
    \caption{
    In each level curve, we present the eigenerror of implementing an $X$ gate by concatenating $C$ pulse shots---each implementing $\sqrt[C]{X}$---using the same total amount of energy (measured in total number of used driving photons, $\navg$). Each shot is achieved using the Jaynes-Cummings interaction, with the driving system initialized in a coherent state with an average of $\navg / C$ photons. Thus, each curve is associated with the drive energy $\navg\hbar\omega$. For the same amount of total energy, using more shots increases on average the channel eigenerror.
    }\label{fig:concatenation_energy_efficiency}
\end{figure}

In Fig.~\ref{fig:concatenation_energy_efficiency} we show numerical results of this analysis done on the $X$ gate, which can be realized through a single JC shot with $\tau=\pi/2$.
We can see that, from an energetic point of view, subdividing a quantum operation increases the channel eigenerror, when the state of the driving system is taken as a Poisson distribution, namely a coherent state.
In Appendix~\ref{appendix:concatenation}, we analytically derive that such an error approximately scales as
\begin{equation} 
    \bar\epsilon_{\theta,C} 
    \approx
    \frac{1}{6 \bar n}
    \left(
        \theta^2
        + C^2 \sin^2\left(\frac{\theta}{C}\right)
    \right).
\end{equation}
In conclusion, the strategy of breaking the target operation in many cycles is not optimal for the \textit{eigenerror-energy balance} when using a fixed amount of total energy in the driving system. Therefore, it appears better to carry out a ``single shot'' evolution that is powered from the beginning by all the available energy, in terms of the number of photons.

%%%%%%%%%%%%%%%%%%%%%%%%%
\section{Conclusions}
%%%%%%%%%%%%%%%%%%%%%%%%%

In this paper, we provide analytical bounds for the error that inevitably arises when one performs a quantum computation by processing the information stored in a quantum logical system via a drive, operated for example by a laser.

First, we introduce from first principles practical quantifiers for the performance of a quantum operation, i.e., the \textit{eigenfidelity} and the \textit{channel eigenfidelity}. They provide upper bounds to the fidelity of the operation in terms of the spectral radius and the purity of the computation output state. From this point of view, these bounds make one understand whether a quantum gate---driven by an external field---is close to its maximum fidelity value, given the characteristics of the drive itself. In particular, the upper bound $(1 + \bar\gamma_t)/2$ of the channel eigenefidelity can be accounted as a \textit{best-case gate fidelity} in using $\mathcal E_t$ to approximate a target unitary logical operation.

Secondly, we show how the eigenfidelities are deeply connected with the energetics of the quantum logical operation of interest.
In accordance with Refs.~\cite{van_enk_classical_2001,banacloche_minimum_energy,banacloche_laser_fields}, we are able to derive the minimal energy requirements for implementing a logic operation within a given target fidelity. Such minimum energy cost arises as a trade-off between two opposing constraints: (i) the entanglement between the drive and the quantum logical system, whose presence can entail a decrease in the logical state's fidelity, and (ii) the minimal time required for the computation, which is bounded by quantum speed limits. In our case study (the Jaynes-Cummings model), we show with analytical and numerical derivations how to balance these constraints in order to attain a minimal fidelity.
Specifically, we compute the scaling of the channel eigenfidelity (representing the quality of the gate that realizes the computation) in performing a given Hilbert-space rotation (space), as a function of the number of photons carried by the drive (energy), the duration of the driving pulse (time), and the number of splitting with which the computation can be achieved (concatenations).
As shown above, not all these requirements can be fulfilled at maximum efficiency, and trade-off conditions have to be established for any quantum gate one aims to implement via a driving system.

As an outlook, it might be interesting to determine whether an exact scaling of the channel eigenfidelity, in relation to the QSLs, holds for even more complex models (in terms of interaction terms) than the Jaynes-Cummings model, and possibly if an experimental validation is feasible.
Nevertheless, these energetic bounds may be improved upon by introducing a further correction system made of ancillary qubits and a supplementary drive, as proposed in \cite{ikonen2017energy}.
We expect that our results can find application to quantum technologies, and especially to quantum computing where linking the fidelity of computation with the needed resources (the time and energy injected by the drive) to achieve it is of fundamental importance for proper engineering of the technology~\cite{AuffevesPRXQuantum2022}. As we go towards working with better and more qubits than ever, the bounds discussed in this paper could become a fundamental challenge at the core of the development of large-scale quantum devices operated by external drives.

\section*{Acknowledgments}
The authors acknowledge Yasser Omar and Sebastian Deffner for the interesting discussions that led to this work. 
S.S.P. acknowledges the support from the ``la Caixa'' foundation through scholarship No. LCF/BQ/DR20/11790030.
S.G. acknowledges support from the PNRR MUR project PE0000023-NQSTI, and the Royal Society IEC\textbackslash R2\textbackslash 222003.

L.B. received funding from Next Generation EU, in the context of the National Recovery and Resilience Plan, M4C2 investment 1.2. Project SOE0000098-ThermoQT. This resource was financed by the Next Generation EU [DD 247 19.08.2022]. The views and opinions expressed are only those of the authors and do not necessarily reflect those of the European Union or the European Commission. Neither the European Union nor the European Commission can be held responsible for them.

\appendix
\section*{Appendices}

%%%%%%%%%%%%%%%%%%%%%%%%%
\section{Evolution of a logical state under the Jaynes-Cummings interaction}\label{appendix:JC_evolution}
%%%%%%%%%%%%%%%%%%%%%%%%%

Let $\mathcal E_t$ be the quantum channel that approximately maps an initial logical state
$\rho(0) = \sum_{i,j} r_{ij} \ketbra{i}{j}$ to another logical state at time $t$, by following the Jaynes-Cummings evolution and tracing out the drive's \textsc{dof}. Since $\mathcal E_t$ is a linear CPTP map, one has that
$\rho(t) = \mathcal E_t[\rho] = \sum_{ij} r_{ij} \, \mathcal E_t[\ketbra{i}{j}]$. Therefore, the evolution of a generic initial logical state can be described by knowing how the operators $\mathcal E_{ij} = \mathcal E_t[\ketbra{i}{j}]$ change over time. The explicit expression of these operators are given by the averages
$\mathcal{E}_{ij} = \mathbb E_\text{dr}[F_{ij}] = \sum_n \abs{b_n}^2 F_{ij}(n)$ over the set of drive coefficients, where the operators $F_{ij}(n)$ are, in the computational basis, represented by
\begin{align*}
    F_{00}(n) &= \begin{pmatrix}
        c_n^2
        & \frac{b_{n+1}}{b_n} \, c_n s_{n+1}
        \\
        \frac{\overline{b_{n+1}}}{\overline{b_n}} \, c_n s_{n+1}
        & s_n^2
    \end{pmatrix} \\
    F_{11}(n) &= \begin{pmatrix}
       s_{n+1}^2
        & -\frac{b_{n+1}}{b_n} \, s_{n+1}  c_{n+2}
        \\
        \frac{\overline{b_{n+1}}}{\overline{b_n}} \, s_{n+1}  c_{n+2}
        & c_{n+1}^2
    \end{pmatrix} \\
    F_{01}(n) &= \begin{pmatrix}
        - \frac{\overline{b_{n+1}}}{\overline{b_n}} c_{n+1} s_{n+1}
        & c_{n+1}  c_{n+2}
        \\
        \frac{b_{n+2}}{b_n} \, s_{n+1}  s_{n+2}
        & \frac{\overline{b_{n+1}}}{\overline{b_n}} c_{n+1} s_{n+1}
    \end{pmatrix} \\
    F_{10}(n) &= {F_{01}(n)}^\dagger,
\end{align*}
with $\omega_n = g\sqrt{n}$, $c_n := \cos(\omega_n t)$, $s_n := \sin(\omega_n t)$, and $\overline z$ is the complex conjugate of $z$. The averages are taken element-wise.

%%%%%%%%%%%%%%%%%%%%%%%%%
\section{Proofs for eigenfidelity of quantum states and channels}\label{appendix:eigenfidelity}
%%%%%%%%%%%%%%%%%%%%%%%%%

\noindent
{\bf Proposition 1.} 
The closest \textit{pure} states to a generic density operator $\varrho$ are the eigenstates (or eigenstate, in the non-degenerate case) $\ket \psi$ corresponding to the largest eigenvalue $r(\varrho)$ of $\varrho$, also denoted as its \textit{spectral radius}. Formally,
\begin{equation}
F(\varrho,\rho_{\varphi}) \leq r(\varrho)
\end{equation}
for any pure state $\rho_{\varphi} := |\varphi\rangle\!\langle\varphi|$, and $F(\varrho,\rho_{\varphi}) = r(\varrho)$ if and only if
$\varrho \ket{\varphi} = r(\varrho) \ket{\varphi}$.

\begin{proof}
    Let $\varrho = \sum_i f_i \ketbra{\psi_i}$ be the spectral decomposition of $\varrho$, with $\sum_{i}f_i = 1$ and $f_1 \geq \cdots \geq f_n$. Thus, the fidelity with which $\varrho$ `approximates' the density  operator $\ketbra{\varphi}$ given by the outer product of a generic state $\ket\varphi$ is
    \begin{equation*}
        \expval{\varrho}{\varphi} 
            = \sum_i f_i\abs{\braket{\psi_i}{\varphi}}^2 
             \leq f_1 \sum_i  \abs{\braket{\psi_i}{\varphi}}^2 = f_1 = r(\varrho).
    \end{equation*}
    Moreover, if $\varrho \ket{\psi} = f_1\ket{\psi}$, then $\expval{\varrho}{\psi} = f_1$ and $|\psi\rangle\!\langle\psi|$ is one of the density operators, given by the outer product of a pure state, that is closest to $\varrho$.
\end{proof}

%%%%%%%%%%%%%%%%%%%%%%%%%
\noindent
{\bf Proposition 2.}
For any $p \geq 1,$ spectral radius of a density operator $\varrho$ is bounded from below and above as    
    \begin{equation}
        \norm{\varrho}_p^q \leq r(\varrho) \leq \norm{\varrho}_p \,,
    \end{equation}
    where $q$ is such that $1/q + 1/p = 1.$
\begin{proof}
The fact that $\norm{\varrho}_p := (f_1^p + \cdots + f_n^p)^{1/p}$ is an upper bound of $r(\varrho)$ is an already known property of the spectral radius. The spectral radius $r(\varrho)$, indeed, is a lower bound of any matrix norm (expressed in terms of $p$) of $\varrho$.
This is a consequence of the Gelfand's formula, which states that 
\begin{equation*}
r(A) = \inf_{p > 0} \norm{A^p}^{1/p} = \lim_{p\rightarrow\infty} \norm{A^p}^{1/p}
\end{equation*}
for any bounded operator $A$ in a Banach space (see, for instance, Example~7.1.4 of Ref.~\cite{meyer2011matrixanalysis} or Theorem 10.13 of \cite{rudin1991functional}).

The lower bound is actually an improvement over the more familiar bound $\norm{\varrho}_p / n^{1/p}.$ In fact, we can prove the following chain of bounds for $p>1$:
\[
    \frac{ \norm{\varrho}_p }{ n^{1/p} } \leq \norm{\varrho}_p^q \leq r(\varrho),
\]
where $q$, also called the \textit{H\"older's conjugate} of $p$, is such that $1/p+1/q=1$, i.e., $q=p/(p-1)$.
Suppose that $f_1 \geq \cdots \geq f_n$ are the eigenvalues of $\varrho$; then, we shall have
\begin{align*}
    \norm{\varrho}^q_p
    & = \left( f_1^p + \cdots + f_n^p \right)^\frac{1}{p-1}\\
    & \leq \left( f_1^{p-1}(f_1+\cdots+f_n) \right)^\frac{1}{p-1}\\
    & = f_1 = r(\varrho).
\end{align*}

We thus need to prove the inequality $\norm{\varrho}_p / n^{1/p} \leq \norm{\varrho}_p^q$. In doing this, first we claim that the maximally mixed state $\tilde \varrho := \mathrm{diag}(1/n, \ldots, 1/n)$ minimizes the $p$-norm over all the density operators, that is,
\[
    \norm{\varrho}_p
    \geq \norm{\tilde \varrho}_p
    = \frac{n^{1/p}}{n}
    \textrm{ for all }\varrho \,.
\]
This result can be proved by using the H\"older's inequality on the vectors of eigenvalues $\underline{f}=(f_1, \ldots, f_n)$ and $\underline{e}=(1/n, \ldots, 1/n)$, leading to
\[
    \norm{\underline{f}\cdot \underline{e}}_1 \leq \norm{ \underline{f} }_p \norm{ \underline{e} }_q
\]
where here $\cdot$ denotes the element-wise product. On the one hand, $\norm{\underline{f}\cdot \underline{e}}_1 = 1/n,$ since $\norm{\underline{f}}_1=1.$ On the other hand, it can be shown that also $\norm{\underline{e}}_p \norm{\underline{e}}_q = 1/n$. Hence,
\[
\norm{\underline{e}}_p \norm{\underline{e}}_q
    = \frac{1}{n}
    = \norm{\underline{f}\cdot \underline{e}}_1
    \leq \norm{\underline{f}}_p \norm{\underline{e}}_q \,.
\]
Dividing both sides by $\norm{\underline{e}}_q$, we get
\[
\norm{\underline{e}}_p \leq \norm{\underline{f}}_p 
\Longleftrightarrow
\norm{\tilde \varrho}_p \leq \norm{\varrho}_p,
\]
with $\norm{\underline{f}}_p$ and $\norm{\underline{e}}_p$ identically equal to $\norm{\varrho}_p$ and $\norm{\tilde \varrho}_p$ respectively. This proves the statement above about the maximally mixed state $\tilde \varrho$. Finally, we carry out the calculation
\[
    \norm{\varrho}_p^q
    =\norm{\varrho}_p \norm{\varrho}_p^{-\frac{1}{(1-p)}}
    \geq \norm{\varrho}_p \norm{\underline{e}}_p^{-\frac{1}{(1-p)}}
    = \frac{\norm{\varrho}_p}{n^{1/p}}
\]
that concludes the proof.
\end{proof}

%%%%%%%%%%%%%%%%%%%%%%%%%
\noindent
{\bf Theorem 1.} The eigenfidelity $r(\varrho)$ and eigenerror $\epsilon(\varrho)$ of $\varrho$ are bounded from below and above as
\begin{equation*}
\gamma \leq r(\varrho) \leq \frac{1 + \gamma}{2}
\quad\text{and}\quad \frac{S_L}{2} \leq \epsilon(\varrho) \leq S_L \,,
\end{equation*}
where $\gamma := \tr \varrho^2$ and $S_L := 1 - \gamma$ denote, respectively, the purity and the linear entropy of $\varrho$.

\begin{proof}
    Let $f_1 \geq \cdots \geq f_n$ be the eigenvalues of $\varrho$ with $f_1 = r(\varrho)$. On the one hand, being $f_i^2 \leq f_1f_i$, we get that
    \[
    \tr \varrho^2 = f_1^2 + \cdots + f_n^2 \leq f_1(f_1 + \cdots + f_n) = f_1 = r(\varrho)\,,
    \]
    i.e., $\gamma \leq r(\varrho)$. On the other hand, using the inequality of arithmetic and geometric means (AM-GM inequality), it holds that
    \[
        \frac{1 + \gamma}{2} = \frac{1 + \tr \varrho^2}{2} \geq \sqrt{1 \cdot \tr \varrho^2} = \sqrt{\tr \varrho^2} \geq r(\varrho)\,,
    \]
    as $0 \leq \gamma \leq 1$. We have thus derived the lower and upper bounds of $r(\varrho)$. Then, the bounds of the eigenerror $\epsilon(\varrho)$ are directly obtained from making the substitution $\epsilon(\varrho) = 1 - r(\varrho)$.
\end{proof}

%%%%%%%%%%%%%%%%%%%%%%%%%
\noindent
{\bf Proposition 3.} For qubits, $\bar \gamma_t$ has a closed-form expression:
\begin{equation*}
\bar \gamma_t = \frac{1}{3} \tr \left(
     \mathcal E_{00}^2
     + \mathcal E_{00} \, \mathcal E_{11}
     + \mathcal E_{11}^2 
     + \mathcal E_{01} \, \mathcal E_{10}
    \right),
\end{equation*}
where, for simplicity, we defined $\mathcal E_{ij} := \mathcal E_t\left[\ketbra{i}{j}\right]$ with $i,j=0,1$.

\begin{proof}
    We have to compute the average value of the purity $\gamma_t := \tr \mathcal{E}_t[\rho]^2$ over all possible pure initial states $\rho$.
    For a qubit, the density operator of a generic initial pure state
    $\ket \alpha
        = e^{ i\varphi}\cos\alpha \ket 0
        + e^{-i\varphi}\sin\alpha \ket 1$
    is represented, in the $\ket 0, \ket 1$ basis as
    \[
        \rho = 
        \frac{1}{2}
        \begin{pmatrix}
            1 + \cos(2\alpha) & e^{i2\varphi}\sin(2\alpha) \\
            e^{-i2\varphi}\sin(2\alpha) & 1 - \cos(2\alpha) 
        \end{pmatrix}.
    \]
    Moreover, the value of $\rho(t) = \mathcal E_t[\rho]$ can be written as a linear combination of the evolved elements $\mathcal E_{ij} := \mathcal E_t\left[\ketbra{i}{j}\right]$ with $i,j=0,1$. Hence, $\gamma_t$ is a function of $\alpha$ and $\varphi$. As a result, taking the average value over $0 \leq \alpha \leq \pi / 2$ and $0 \leq \varphi \leq \pi$ yields the desired result, which was checked using Sagemath~\cite{sagemath}.
\end{proof}

%%%%%%%%%%%%%%%%%%%%%%%%%
\section{Bounds on quantum gates fidelity}
%%%%%%%%%%%%%%%%%%%%%%%%%

The results in Section \ref{section:entanglement_induced_error} of the main text can be specialized to provide lower and upper bounds to the fidelity of a quantum gate that we wish to implement. In doing this, we take the quantum gate $U$ that brings $|\alpha\rangle$ to $|\delta\rangle$, i.e., $\ket\delta = U \ket\alpha$, so that Eq.~\eqref{eq:QSL-general} in the main text becomes
\begin{align}\label{eq:QSL-target-gate}
    F(\mathcal E_t[\rhoinput], U\rhoinput U^\dagger)
       & = \expval{ U^\dagger \mathcal E_t \left(\ketbra{\alpha}\right) U}{\alpha}\nonumber \\
       & = \tr \left[ \left(\ketbra{\alpha}^T \otimes \ketbra{\alpha}\right) S_{U,t}\right], 
\end{align}
with $S_{U, t}$ denoting the Choi matrix~\cite{wolf2012quantum} associated to the transformation $\rho_{\alpha} \to U^\dagger \mathcal E_t[\rho_{\alpha}] U$. Using Eq.~(\ref{eq:QSL-target-gate}) and the linearity of the trace, one can compute the average fidelity associated with the quantum gate $U$, i.e., 
\begin{equation}\label{eq:average_fidelity}
 \bar F(\mathcal E_t,U)
    := \int \dd \mu(\rho_{\alpha})
        F(\mathcal E_t[\rho_{\alpha}], U\rho_{\alpha} U^\dagger)
    = \tr (A S_{U, t})
\end{equation}
where $A$ is a $d^2 \times d^2$ matrix obtained by averaging $\rho_{\alpha}^T \otimes \rho_{\alpha}$ over all the pure states $\rho_{\alpha}$. For example, for a logical state encoded in a qubit,
\[
A = 
\frac 1 6
\left(
\begin{smallmatrix}
    2 & 0 & 0 & 1 \\
    0 & 1 & 0 & 0 \\
    0 & 0 & 1 & 0 \\
    1 & 0 & 0 & 2
\end{smallmatrix}
\right).
\]
The value $\bar F(\mathcal E_t,U)$ is the average fidelity with which the quantum channel $\mathcal E_t$ ``approximates'' the quantum gate $U$. Eq.~(\ref{eq:average_fidelity}) equals to the \textit{average quantum gate fidelity}, Eq.~(18), in \cite{Nielsen2002}.

Combining Proposition \ref{prop:closest_state_fidelity} and Corollary~\ref{cor:avg_eigenfidelity_bounds} gives us an upper bound also on $\bar F(\mathcal E_t, U)$. Such a bound is provided by the channel eigenfidelity $\bar r(\mathcal E_t)$, which in turn is bounded from above by $(1 + \bar \gamma)/2$. Hence, for any time $t$,
\begin{equation*}
    \bar F(\mathcal E_t, U) \leq \bar r(\mathcal E_t) \leq \frac{ 1 + \bar \gamma_t }{2} \,.
\end{equation*}

%%%%%%%%%%%%%%%%%%%%%%%%%
\section{Bounds on the concatenation of quantum gates}
\label{appendix:concatenation}
%%%%%%%%%%%%%%%%%%%%%%%%%

For a qubit, any quantum channel $\mathcal E$ can be represented by the Pauli Transfer Matrix (PTM) $P \in \mathbb{C}^{4\times 4}$ given by~\cite{guo2015gate}
\[
P=\begin{pmatrix}
    1 & \underline{0} \\
    \underline{v} & A
  \end{pmatrix},
\]
whose elements are $P_{ij} \equiv \tr(\sigma_i \, \mathcal E[\sigma_j])$, with $\sigma_i$ denoting---as usual---the $2\times 2$ Pauli matrices including the identity. Moreover, $A$ is a $3 \times 3$ matrix, $\underline{v}$ is a column vector of dimension $3$, and $\underline{0}$ is a row vector $\in\mathbb{R}$ of zeros.

In the PTM formalism, a quantum state $\rho$ is represented by the Bloch vector
\[
\underline{b} \equiv \left[ \tr(\rho\, \sigma_1), \tr(\rho\, \sigma_2), \tr(\rho\, \sigma_3) \right] \,,    
\]
and the eigenerror is returned by
\[
    \epsilon(\rho) = \frac{1 - \norm{ \underline{b} }}{2}\,.
\]
Then, the quantum state $\mathcal E[\rho]$ is represented by the vector
\[
    \underline{b}_1 = A\underline{b}_0 + \underline{v}\,.
\]

Repeated application of the quantum channel $\mathcal E$ leads the system to converge to a fixed-point, which we can calculate in the PTM formalism as $\underline{b}^{*} = A\underline{b}^{*} + \underline{v}$, or
\[
    \underline{b}^{*} = (\mathbb{I}-A)^{-1}\underline{v}\,.
\]
Notice that it is implicitly required that $(\mathbb{I}-A)$ is invertible, which is true for the most physical cases. Otherwise, to solve the equation above, the Moore-Penrose pseudo-inverse can be used. In terms of this fixed-point, concatenating $C$ times the channel $\mathcal{E}$ on the quantum state $\rho$ results in the Bloch vector
\begin{align*}
    \underline{b}_C &= A^{C}\underline{b}_{0} + A\left( \mathbb{I} - A^{C-1} \right)\underline{b}^{*} + \underline{v}\\
    &= A^{C}\underline{b}_{0} + (\mathbb{I} - A^C)\underline{b}^{*},
\end{align*}
which indeed approaches $\underline{b}^{*}$ if $\norm{A}_\infty < 1$.

Our goal is to calculate the average of the channel eigenerror
\[
\frac{1 - \norm{\underline b_C}}{2}
\]
over all input pure states $\underline b_0$. For this purpose, we first make the approximation $\underline b^* \approx 0$ (such that $\underline b_C \approx A^C \underline b_0$), which is valid for short interaction times $\tau \leq \pi$. Then, since calculating the average of $\norm{A^C \underline b_0}$ (i.e., $\overline{\norm{A^C \underline b_0}}$) is a hard task, we approximate $\overline{\norm{A^C \underline b_0}}$ through its root-mean-square:
\begin{align}\label{eq:approx_average_norm_power-A}
\int\,\dd \mu(\underline b_0)
\norm{A^C \underline b_0}
& \approx
\sqrt{
\int \dd \mu(\underline b_0)
\norm{A^C \underline b_0}^2
}\nonumber \\
& = \sqrt{
\frac{
    \sigma_1^2(A^C)
    + \sigma_2^2(A^C)
    + \sigma_3^2(A^C)
    }{3}
}\,,
\end{align}
where $\sigma_1(A^C) \leq \sigma_2(A^C) \leq \sigma_3(A^C)$ are the singular values of $A^C$. It is worth noting that, albeit convenient for computation purposes, the approximation in the first line of Eq.~(\ref{eq:approx_average_norm_power-A}) is clearly not valid in general. Thus, before proceeding, we made sure that in the parameters regime of interest for our case study involving the Jaynes-Cummings model, the approximation in (\ref{eq:approx_average_norm_power-A}) is fulfilled with negligible error.

At this point, in the case Eq.~(\ref{eq:approx_average_norm_power-A}) is approximately valid, one needs to determine the singular values of $A^C$, whose analytical expression is difficult to determine in the general case. We thus resort to the Gelfand's formula whereby $\sigma_3(A^C)^{1/C} \longrightarrow \abs{\lambda_3(A)}$, in the limit of $C$ large ($C\to\infty$), where $\lambda_3(A)$ refers to the largest eigenvalue (in absolute value) of $A$. In the case where the logical qubit is operated by the Jaynes-Cummings Hamiltonian, $A$ has a particular block structure (a $1\times 1$ and a $2\times 2$ blocks, where the two eigenvalues of the $2\times 2$ block have the same modulus, as shown below) that makes the Gelfand's formula valid for all the three singular values, such that $\sigma_i(A^C) \approx \abs{\lambda_i(A)}^C$.
\begin{figure}[b!]
    \centering
    \resizebox{0.95\columnwidth}{!}{\includegraphics{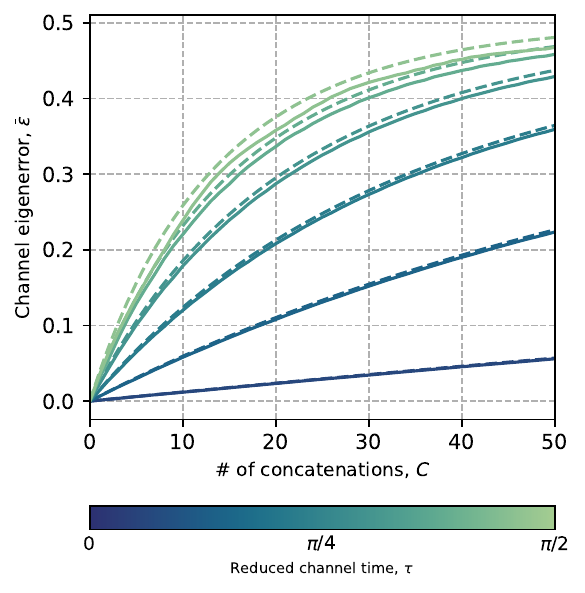}}
    \caption{
    Channel eigenerror committed from concatenating $C$ identical logical operations operated by the Jaynes-Cummings evolution. In this figure, the initial state of the drive is a binomial distribution with $\navg=25$, $s=\nstdev^2 / \navg=1/2$, and the reduced interaction time $\tau$ is made vary in the interval $[0,\pi/2]$. Solid line: numerical simulation; Dashed line: analytical approximation of the channel eigenerror as provided by Eq.~\eqref{eq:concat-eigenerror-approx}.}
    \label{fig:concat-eigenerror}
\end{figure}
\begin{figure}[t!]
    \centering
    \includegraphics[width=1\linewidth]{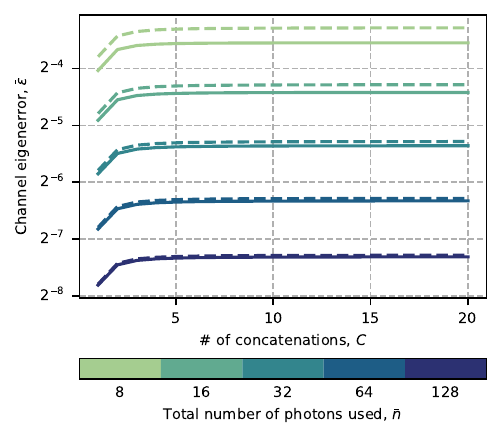}
    \caption{
        Comparison between the analytical (dashed lines) and numerical (solid lines) scaling of the channel eigenerror in concatenating $C$ logical operations enabled by the Jaynes-Cummings evolution. The numerical scaling is exactly the one in Fig.~\ref{fig:concatenation_energy_efficiency}, while the analytical prediction of the scaling is provided by Eq.~\eqref{eq:concatenation-energy-efficiency-approximation}.
    }
    \label{fig:concatenation-energy-efficiency-bounded}
\end{figure}
If we assume that the drive is initialized in a state with a binomial distribution, then the eigenvalues of the matrix $A$ have just two distinct values in modulus:
\begin{align*}
    \abs{\lambda_{a}}
    = \abs{\lambda_b}
    & \approx \sqrt{
        1 - \frac{1}{\navg}
        \left(s\tau^2 + \frac{\sin^{2}(\tau)}{2s}\right)
    } \\
    \abs{\lambda_c}
    & \approx \sqrt{
        1 - \frac{\sin^{2}(\tau)}{s \navg}
    }\,.
\end{align*}
Depending on $s$, either eigenvalue may be dominant. As a result, using these values and the approximations above, the average channel eigenerror is approximately equal to
\begin{equation}
\label{eq:analytical_eigenerror_k}
\bar \epsilon_{\tau, C}
\approx
\frac{1}{2}
- \frac{1}{2}
\sqrt{
\frac{
    \abs{\lambda_1}^{2C}
    + \abs{\lambda_2}^{2C}
    + \abs{\lambda_3}^{2C}
    }{3}
    }
\end{equation}
where, we recall, by definition $\abs{\lambda_1} \leq \abs{\lambda_2} \leq \abs{\lambda_3}$. 
Under the additional assumption that the value of the dominant eigenvalue $\lambda_3$ is, in modulus, larger than the other eigenvalues, then \eqref{eq:analytical_eigenerror_k} can be further approximated as
\begin{equation}\label{eq:analytical_eigenerror_lb}
\bar \epsilon_{\tau, C} 
\approx \frac{1 - \abs{\lambda_3}^C}{2}
\end{equation}
that we have used to explain the behaviours shown in Fig.~\ref{fig:concatenations} with theoretical arguments. Specifically, for $\navg=25$ and $s=1/2$ (as in Fig.~\ref{fig:concatenations}), (\ref{eq:analytical_eigenerror_lb}) becomes
\begin{equation}\label{eq:concat-eigenerror-approx}
    \bar\epsilon_{\tau,C} \approx
    \frac{1}{2}
    - \frac{1}{2}
    \left( 1 - \frac{s \tau^2 +\sin^2(\tau)}{2s\navg} \right)^{C/2} \,.
\end{equation}
In Fig.~\ref{fig:concat-eigenerror} we plot the comparison between the approximated analytical value of $\bar\epsilon_{\tau,C}$ as given in Eqs.~(\ref{eq:analytical_eigenerror_k}) and the channel eigenerror obtained numerically using for the drive state a binomial distribution with $\navg=25$ and $s=1/2$.

In the same way, we can also compare the channel eigenerror of an evolution with duration $\tau$ employing $\bar n$ photons, against $C$ evolutions of time $\tau / C$ and $\bar n / C$ photons. For the case studied in Fig.~\ref{fig:concatenation_energy_efficiency} of the main text, the dominant eigenvalue (in modulus) is $\lambda_c$ such that the corresponding channel eigenerror follows the approximate behaviour
\begin{equation}
    \bar\epsilon_{\tau,C} 
    \approx
    \frac{1}{6 \bar n}
    \left(
        \tau^2
        + C^2 \sin^2\left(\frac{\tau}{C}\right)
    \right)\,.
    \label{eq:concatenation-energy-efficiency-approximation}
\end{equation}
(\ref{eq:concatenation-energy-efficiency-approximation}) well approximates the behaviours observed in Fig.~\ref{fig:concatenation_energy_efficiency}, as illustrated in Fig.~\ref{fig:concatenation-energy-efficiency-bounded}.

%%%%%%%%%%%%%%%%%%%%%%%%%

\bibliographystyle{unsrt}
\bibliography{references}

\end{document}